\documentclass[12pt,english]{article}
\usepackage{babel}
\usepackage{amsfonts}
\title{NJL model derived from QCD
}
\author{B.A. Arbuzov$^1$
, M.K. Volkov$^2$ and I.V. 
Zaitsev$^3$\\
{\it $^1$Skobeltsyn Institute of Nuclear Physics of
MSU,}\\ {\it 119992 Moscow, RF},\\
e-mail: arbuzov@theory.sinp.msu.ru\\
{\it $^2$Joint Institute of Nuclear Research,}\\
{\it 141980 Dubna, Moscow Region, RF},
\\
{\it $^3$Physics Department of MSU, 119992 Moscow, RF}}

\date{}
\newcommand{\be}{\begin{equation}}
\newcommand{\ee}{\end{equation}}
\newcommand{\beq}{\begin{eqnarray}}
\newcommand{\eeq}{\end{eqnarray}}
\newcommand{\nn}{\nonumber}
\newcommand{\bi}{\bibitem}
\begin{document}
\maketitle
\begin{quote}

We apply Bogolubov approach to QCD with two light quarks to
demonstrate a  spontaneous generation of an effective interaction,
leading to the Nambu -- Jona-Lasinio model. The resulting theory
contains two parameters: average low-energy value of $\alpha_s$ and
current light quark mass $m_0$. All other low-energy parameters: the
pion decay constant, mass of the $\pi$-meson, mass of the
$\sigma$-meson and its width, the constituent quark mass, the quark
condensate are expressed in terms of the two input parameters in
satisfactory correspondence to experimental data and chiral
phenomenology. {\it E.g.} in the approximation being used we have for
$\alpha_s = 0.67$ and $m_0 = 20\, MeV$: $f_\pi = 93\,MeV,\, m_\pi =
135\,MeV,\, m_\sigma = 492\,MeV,\, \Gamma_\sigma = 574\,MeV, \,m_q =
295\,MeV,\,<\bar q\,q> = -(222\,MeV)^3$.
\end{quote}


\section{Introduction}

The Nambu -- Jona-Lasinio model~\cite{Nambu, Jap, Vol, Vol2} has 
manifested itself to be a good phenomenological tool for low-energy 
hadron
physics.
It is well-known, that the fundamental
perturbative theory QCD is valid in region of large~$q^2$. 
In low-momenta region NJL model supplements the fundamental
QCD.  It is important, that common property of
both theories consists in the chiral symmetry, which defines main
features of low-energy hadron physics. However till now there was no
direct derivation of NJL model from QCD. Therefore the problem to find a
relation between parameters of NJL and those of QCD for a long time
was quite actual. Some attempts in this direction were accompanied by
inevitable introduction of additional parameters (see, {\it
e.g.}~\cite{ERV, SimNJL}).

In work~\cite{Arb05}  with  the use of results~\cite{Arb04} based
on the method developed by N.N. Bogolubov~\cite{Bog2, Bog} an approach 
is proposed
to obtain an effective interaction of the Nambu -- Jona-Lasinio
model, which contains no additional parameters but the QCD ones. Main
parameters of low-energy hadron physics were calculated
there~\cite{Arb05} in chiral limit that is for current mass
of light quarks $m_0 = 0$. In the present work we apply the approach
for description of light zero-spin mesons with account of non-zero
current quark mass.

The structure of the work is the following.\\
In Section 2 the spontaneous generation of NJL type interaction in
QCD is described.\\
Section 3 deals with the description of the scalar and pseudo-scalar states (mesons).\\
In Section 4 spontaneous violation of the chiral invariance is 
demonstrated.\\
In Section 5 expressions for quark condensate and for parameters of
zero-spin mesons are presented.\\
Section 6 is devoted to numerical results and discussions.

\section{Effective NJL interaction}

Now we start with QCD Lagrangian with two light quarks
($u$ and $d$) with number of colours $N = 3$
\be
L\,=\,\sum_{k=1}^2\biggl(\frac{\imath}{2}
\Bigl(\bar\psi_k\gamma_
\mu\partial_\mu\psi_k\,-\partial_\mu\bar\psi_k
\gamma_\mu\psi
_k\,\biggr)-\,m_0\bar\psi_k\psi_k\,+
\,g_s\bar\psi_k\gamma_\mu t^a A_\mu^a\psi_k
\biggr)\,-
\,\frac{1}{4}\,\biggl( F_{\mu\nu}^aF_{\mu\nu}^a\biggr);
\label{initial}
\ee
where we use the standard QCD notations.

Let us assume that a non-local NJL interaction is spontaneously 
generated in this theory. We use Bogolubov approach~\cite{Bog2, Bog} 
to check this assumption. 
In accordance to the approach, application of which to such
problems are described in details in work~\cite{Arb04}, we look
for a non-trivial solution of a compensation equation, which is
formulated on the basis of the Bogolubov procedure {\bf add --
subtract}. Namely let us rewrite the initial
expression~(\ref{initial}) in the form
\beq
& &L\,=\,L_0\,+\,L_{int}\,;\nn\\
& &L_0\,=\,\frac{\imath}{2}
\Bigl(\bar\psi\gamma_
\mu\partial_\mu\psi-\partial_\mu\bar\psi\gamma_\mu\psi
\biggr)\,-\,\frac{1}{4}\,F_{0\,\mu\nu}^aF_{0\,\mu\nu}^a\,-
\,m_0\bar\psi\,\psi\,+
\,\frac{G_1}{2}\cdot\Bigl(
\bar\psi\tau^b\gamma_5\psi\,\bar\psi \tau^b\gamma_5
\psi\,-\nn\\
& &-\bar\psi\,\psi\,\bar\psi\,\psi\biggr)\,+\,\frac{G_2}
{2}\cdot\Bigl(\bar\psi\tau^b\gamma_\mu\psi\,\bar\psi
\tau^b\gamma_\mu\psi +
\bar\psi \tau^b\gamma_5 \gamma_\mu \psi \bar\psi \tau^b
\gamma_5 \gamma_\mu \psi\biggr)\,;\label{addsub}\\
& &L_{int}\,=\,g_s\,\bar\psi\gamma_\mu
t^a A_\mu^a\psi\,-\,\frac{1}{4}\,\biggl( F_{\mu\nu}^aF_
{\mu\nu}^a -
F_{0\,\mu\nu}^aF_{0\,\mu\nu}^a\biggr)\,-\,\frac{G_1}{2}
\cdot\Bigl(\bar\psi \tau^b\gamma_5 \psi\,\bar\psi \tau^b
\gamma_5\psi -\bar\psi\,\psi\,\bar\psi\,\psi\biggr)\,-
\nn\\
& &-\,\frac{G_2}{2}\cdot\Bigl(\bar\psi \tau^b\gamma_\mu
\psi\,\bar\psi \tau^b\gamma_\mu \psi +
\bar\psi \tau^b\gamma_5 \gamma_\mu \psi \bar\psi \tau^b
\gamma_5 \gamma_\mu \psi\biggr)\,.\label{intas}
\eeq
Here $\psi$ is the isotopic doublet of quark fields, colour summation is
performed inside of each fermion bilinear combination, $F_{0\,\mu\nu} =
\partial_\mu A_\nu - \partial_\nu A_\mu$, and notation
 $G_1\cdot \bar\psi \psi \bar\psi \psi$ corresponds to
non-local vertex in the momentum space
\be
\imath\,(2\pi)^4\,G_1\,F_1(p1,p2,p3,p4)\,
\delta(p1+p2+p3+p4)\,;
\label{vertex}
\ee
where $F_1(p1,p2,p3,p4)$ is a form-factor and
$p1,\;p2,\;p3,\;p4$ are incoming momenta.
In the same way we define vertices, containing
Dirac and isotopic matrices. We comment the composition
of the vector sector, which here contain only isovector terms,
in what follows.

Let us consider  expression~
(\ref{addsub}) as the new {\bf free} Lagrangian $L_0$,
whereas expression~(\ref{intas}) as the new
{\bf interaction} Lagrangian $L_{int}$. Then
compensation conditions (see again~\cite{Arb04}) will
consist in demand of full connected four-fermion vertices,
following from Lagrangian $L_0$, to be zero. This demand
gives a set of non-linear equations for form-factors
$F_i$.

These equations according to terminology of 
works~\cite{Bog2, Bog} are called {\bf compensation equations}.
In a study of these equations the
existence of a perturbative trivial solution (in our case
$G_i = 0$) is always evident, but a non-perturbative
non-trivial solution may also exist. Just the quest of
a non-trivial solution inspires the main interest in such
problems. It is impossible to find an exact
non-trivial solution in a realistic theory, therefore
the goal of a study is a quest of an adequate
approach, the first non-perturbative approximation of
which describes the main features of the problem.
Improvement of a precision of results is to be achieved
by corrections to the initial first approximation.

Thus our task is to formulate the first approximation.
Here the experience acquired in the course of performing
of work~\cite{Arb04} is useful. Now in view of
obtaining the first approximation we would make the following
assumptions.\\
1) In compensation equations we restrict ourselves by
terms with loop numbers 0, 1, 2. For one-loop case only trivial 
solution exists. Two-loop terms lead to integral equations, 
which may have non-trivial solutions. So the account of two-loop terms 
leads to the first non-trivial approximation.\\
2) In compensation equations we perform a procedure of
linearizing over form-factor, which leads to linear integral
equations. It means that in loop terms only one vertex contains the
form-factor, while other vertices are considered
to be point-like. In diagram form equation for form-factor $F_1$ is
presented in fig. 1. An accuracy of this procedure was estimated in 
work~\cite{Arb04} to be of order of magnitude of few {\it per cent}.\\
3) While evaluating diagrams with point-like vertices diverging
integrals appear. Bearing in mind that as a result of the study we
obtain form-factors decreasing at momentum infinity, we use an 
intermediate regularization by introducing UV cut-off $\Lambda$ in the 
diverging integrals. It will be shown that results do not depend on the 
value of this cut-off.\\
4) We use a special approximation for integrals, which is connected with 
transfer of a quark mass from its propagator to the lower limit of 
momentum integration. Effectively this leads to introduction 
of IR cut-off at the lower limit of
integration by Euclidean momentum squared $q^2$ at value $m^2$. 
To justify this prescription let us consider a typical integral 
to be encountered here and perform simple evaluations. Functions which 
we use here depend on variable of the form $\alpha\,q^2$, where $\alpha$ 
is a parameter having $1/m^2$ dimension. Now we have  
\beq
& &\int_0^\infty \frac{F(\alpha\,q^2)\,dq^2}{(q^2 + m^2)^k}\,=\,\int_{m^2}
^\infty \frac{F(\alpha\,(q'^2 - m^2))\,dq'^2}{(q'^2)^k}\,=\nn\\
& &=\,\int_{m^2}^\infty 
\frac{F(\alpha\,q'^2)\,dq'^2}{(q'^2)^k}\,-\,\alpha\,m^2\,\int_{m^2}^\infty 
\frac{F'(\alpha\,q'^2)\,dq'^2}{(q'^2)^k}\,+\,...\,;\label{approx}
\eeq
Thus the introduction of IR cut-off at $m^2$ 
which actually consists in the following substitution
\be
\int_0^\infty \frac{F(\alpha\,q^2)\,dq^2}{(q^2 + m^2)^k}\,\to\,\int_{m^2}^
\infty \frac{F(\alpha\,q^2)\,dq^2}{(q^2)^k}\,;\label{subst}
\ee
corresponds to accuracy, which is defined by parameter $\alpha\,m^2$. As 
we shall see, this parameter for our solutions does not exceed order of 
magnitude of few {\it per cent}. 
We use this tool throughout the present work. In doing 
this we keep at nominators only the leading terms in
$m$ expansions because taking into account of the next terms
evidently means supererogation of accuracy. Note, that similar method
was used in work~\cite{VYu} for description of the confinement in
NJL model.\\
5) We shall take into account only the first two
terms of the $1/N$ expansion. Neglected terms gives contribution, 
which values are defined by parameter $1/(4\,N)$. Here additional 
factor $4$ in the denominator is connected with structure of NJL 
interaction in Lagrangian~(\ref{intas}). Indeed a trace in colour indices  
is always accompanied by a trace in spinor indices, which gives 
factor $4$. Thus this approximation  defines accuracy 
$\simeq\,8\,\%$.

Let us formulate compensation equations taking into account all the 
introduced prescriptions.
For {\bf free} Lagrangian $L_0$~(\ref{addsub}) full connected
four-fermion vertices are to vanish. One can succeed in
obtaining analytic solutions for the following set
of momentum variables (see fig. 1): left-hand legs
have momenta  $p$ and $-p$, and right-hand legs
have zero momenta. In particular this kinematics
suits for description of zero-mass bound states.
The construction of expressions with
an arbitrary set of momenta is the problem for the
subsequent approximations. In the present work we shall
use the next approximation for obtaining parameters of scalar and 
pseudo-scalar mesons.

Now following the rules being stated above we
obtain the following equation for form-factor $F_1(p)$ in
scalar channel
\beq
& &G_1 F_1(p^2) =
\frac{G_1^2 N \Lambda^2}{2 \pi^2}\biggl(1 + \frac{1}{4 N}
-\frac{G_1 N }{2 \pi^4}\Bigl(1 + \frac{1}{2 N}\Bigr)
\int \frac{F_1(q^2)\,dq}{q^2}\,\biggr)+
\nn\\
& &+\,\frac{3 G_1\, G_2}{8 \pi^2}\biggl
(2 \,\Lambda^2 + p^2 \log\frac{p^2}
{\Lambda^2}-\frac{3}{2}\, p^2 - \frac{m_0^4}{2\,p^2}
\biggr)\, -\,\frac{(G_1^2+6 G_1 G_2) N}{32\, \pi^6}
\times\nn\\
& & \times \int\Bigl( 2 \Lambda^2+(p-q)^2
\log\frac{(p-q)^2}{\Lambda^2}-\frac{3} {2} (p-q)^2 - \frac{m_0^4}{2
(p-q)^2} \biggr) \frac{ G_1 F_1(q^2) dq}{q^2}\, .\label{eq1} \eeq
Here integration is performed in the four-dimensional Euclidean
momentum space with IR cut-off at $m_0^2$. One-loop expressions contains 
terms
proportional to $N$ and $1$ while two-loop terms correspond to $N^2$
and $N$. The leading terms are the same for scalar and pseudo-scalar
cases. We perform the study with the scalar channel, because it
defines spontaneous chiral symmetry breaking effect.
Equation~(\ref{eq1}) evidently has trivial solution $G_1 = 0$.
Bearing in mind our goal to look for non-trivial solutions we divide
the equation by $G_1$ and perform angular integration in
four-dimensional Euclidean space. From~(\ref{eq1}) we have
\beq &
&F_1(x) = A + \frac{3\,G_2}{8 \pi^2}\biggl (2\Lambda^2 + x
\log\frac{x} {\Lambda^2}-\frac{3}{2}\, x -\frac{\mu^2}{2 x}\biggr) -
\frac{(G_1^2+6 G_1 G_2) N}{32\, \pi^4}\times\nn\\
& &\times \Biggl(\frac{1}{6\,x}
\int_\mu^x (y^2-3 \mu^2) F_1(y)\,dy\,+\,\frac{3}{2}
\int_\mu^x y F_1(y)\,dy +
\log\,x \int_\mu^x y F_1(y)\,dy +\nn\\
& &+ x\,\log\,x \int_\mu^x
F_1(y)\,dy +
\int_x^\infty y \log\,y\, F_1(y)\,dy +\,x \int_x^\infty
\Bigl(\log\,y\,+\frac{3}{2}\biggr) F_1(y)\,dy\,+\nn\\
& &+\,\frac{x^2-3 \mu^2}{6}
\int_x^\infty\,\frac{F_1(y)}{y}\,dy + \Bigl(2\Lambda^2-
\frac{3}{2}\, x\Bigr)\int_\mu^\infty F_1(y)\,dy -
\frac{3}{2}\,\int_\mu^\infty y F_1(y)\,dy\,-\nn\\
& &-\,\log\,
\Lambda^2 \Bigl(\int_\mu^\infty y F_1(y)\,dy\,+\,
x \int_\mu^\infty F_1(y)\,dy\Bigr)\,\Biggr)\,;\label{eq1xy}
\\
& &A\, = \,
\frac{G_1^2 N \Lambda^2}{2 \pi^2}\biggl(1 + \frac{1}{4 N}
-\frac{G_1 N }{2 \pi^2}\Bigl(1 + \frac{1}{2 N}\Bigr)
\int_\mu^\infty F_1(y)\,dy\,\biggr)\,;\nn\\
& &\mu\,=\,m_0^2\,;\qquad x\,=\,p^2\,;\qquad y\,=\,q^2\,.\nn
\eeq
Equation~(\ref{eq1xy}) by a sequential 
six-fold differentiation reduces to the following 
differential equation
\beq
& &\frac{d^2}{dx^2}\Biggl(x\,\frac{d^2}{dx^2}\biggl(x\,
\frac{d^2}{dx^2}\Bigl(x\,F_1(x)\biggr)+\frac{\beta\,
m_0^4}{4}\,F_1(x)\biggr)\Biggr)\,=\,
\beta\,\frac{F_1(x)}{x}\,.\label{diff1}\\
& &\beta\,=\,\frac{(G_1^2+6\,G_1 G_2)\,N}{16\,\pi^4}\,;
\nn
\eeq
with boundary conditions to be formulated below.

Equation~(\ref{diff1}) reduces to Meijer 
equation~\cite{be}. Namely with the simple 
substitution we have
\beq
& &\Biggl(\biggl(z\,\frac{d}{dz}-b\biggr)\biggl(z\,\frac
{d}{dz}-a\biggr)\,z\,\frac{d}{dz}\biggl(z\,\frac{d}{dz}-
\frac
{1}{2}\biggr)\biggl(z\,\frac{d}{dz}-\frac{1}{2}\biggr)
\biggl
(z\,\frac{d}{dz}-1\biggr) - z\Biggr)
 F_1(z) =\,0\,;\label{Meijer}\\
& &z\,=\,\frac{\beta\,x^2}{2^6}\,;\quad a\,=\,-\,\frac{1-
\sqrt{1- 64 u_0}}{4}\,;\quad 
b\,=\,-\,\frac{1+\sqrt{1- 64 u_0}}{4}\,;\quad 
u_0\,=\,\frac{\beta\,m_0^4}{64}\,.\nn
\eeq
Boundary conditions for equation~(\ref{Meijer}) are 
formulated in the same way as in works~\cite{Arb05}, 
\cite{Arb04}. At first we have to choose solutions 
decreasing at infinity, that is combination of the 
following three solutions 
\beq
& &F_1(z)\,=\,C_1\,G_{06}^{40}\Bigl(z\,
|1,\,\frac{1}{2},\,\frac{1}{2},\,0,\,a,\,
b\Bigr)\,+\,C_2\,G_{06}^{40}\Bigl(z\,|1,\,\frac{1}{2},\,b,
\,a,\,\frac{1}{2},\,0,\,\Bigr)\,+\nn\\
& &+\,C_3\,G_{06}^{40}\Bigl(z\,
|1,\,0,\,
b,\,a,\,\frac{1}{2},\,\frac{1}{2}\Bigr)\,;\label{Ci}
\eeq
where
$$
G^{m n}_{p q}\biggl( z\,|^{a_1,\,...\,,\,a_p}_
{b_1,\,...\,,\,b_q}\biggr)\,;
$$
is a Meijer function~\cite{be} with
sets of upper indices ${a_i}$ and of lower ones ${b_j}$.
In case only one line of parameters is written this
means the presence of lower indices only, $n$ and $p$ in
the case being equal to zero.

Constants $C_i$ are defined by boundary conditions
\be
\frac{3\,G_2}{8 \pi^2}-\frac{\beta}{2}
\int_{m^2_0}^\infty F_1(y)\,dy\,=\,0\,;\quad
\int_{m^2_0}^\infty y\,F_1(y)\,dy\,=\,0\,;\quad
\int_{m^2_0}^\infty y^2\,F_1(y)\,dy\,=\,0\,;\label{bound}
\ee
which one obtains from integral equation~(\ref{eq1xy}) by 
considering asymptotic behaviour of integral terms at infinity.  
These conditions and condition $A = 0$ as well provide
cancellation of all terms in equation~(\ref{eq1xy}) being
proportional to $\Lambda^2$ and $\log\,\Lambda^2$. Thus
the result does not depend on a value of parameter
$\Lambda$.
By solving linear set~(\ref{bound}), in which
solution~(\ref{Ci}) is substituted, we obtain the unique
solution. Value of parameter $u_0$, which is connected
with initial quark mass, and ratio of two constants $G_i$
we obtain from conditions $F_1(\mu)=1$ and
\beq
& &A\,=\,\frac{G_1 N \Lambda^2}{2 \pi^2}\biggl(1+ \frac{1}
{4 N} -\,\frac{G_1 N }{2 \pi^2}\Bigl(1 + \frac{1}{2 N}
\Bigr)\int_{m^2_0}^\infty F_1(y)\,dy\,\biggr)\,=\nn\\
& &=\,\Bigl(1+ \frac{1}
{4 N}\Bigr)\frac{G_1 N \Lambda^2}{2 \pi^2}\biggl(1- \frac{6 G_2 (4 N+2)}
{(G_1 + 6 G_2)(4 N+1)}\biggr)\,=\,0\,;\label{A0}\\
& &F_1(u_0)\,=\,1\,;\qquad u_0\,=\,\frac{\beta\,m_0^4}{2^6}\,=\,
\frac{N\,(G_1^2+6 G_1 G_2)\,m_0^4}{1024 \,\pi^4}\,.\nn
\eeq
The last line here presents the obvious condition of normalization of 
a form-factor on the mass shell.  
Now relations~(\ref{A0}) give for $N=3$ with the account of the first of
conditions~(\ref{bound})
\be
u_0\,=\,1.92\cdot
10^{-8}\,\simeq 2\cdot 10^{-8}\,;\quad
G_1\,=\,\frac{6}{13}\,G_2\,.\label{result1}
\ee
So $G_1$ and $G_2$ are both defined in terms of $m_0$. Thus we have the
unique non-trivial solution of the compensation equation, which contains
no additional parameters.

The form-factor now reads as~(\ref{Ci}) with
\be
C_1\,=\,0.28\,;\quad C_2\,=\,-\,3.66\cdot10^{-8}\,;
\quad C_3\,=\,-\,7.8\cdot10^{-8}\,;\label{Cir}
\ee
In what follows we use the notation $F_1(z)$ for expression~(\ref{Ci}), 
where $z$ is always the dimensionless variable defined 
in eq.(\ref{Meijer}). 
We have 
$F_1(u_0) = 1$ and $F_1(z)$ decreases with $z$ increasing in the 
following way
$$
F_1(z) \to \frac{D}{z^\frac{1}{6}}\,\exp\bigl(-\,3(1 - \imath 
\sqrt{3})\,z^\frac{1}{6}\bigr)\,+\,h.c.\,;
$$
where $D$ is a complex constant. 
It is important, that the solution exists only for
positive $G_2$ and due to~(\ref{result1}) for positive
$G_1$ as well.

At this point we would comment the problem of accuracy of 
our method of taking into account of quark mass $m_0$. 
A possible corrections being proportional to $m_0^2$ correspond 
to dimensionless variable~(\ref{approx}) where $\alpha = \sqrt{\beta}/8$
\be
\alpha\,m_0^2\,=\,\sqrt{u_0}\,\simeq\,10^{-4}\,;\label{acm0}
\ee
and so they are not significant for definition of form-factor $F_1(z)$.
 
We would also comment the composition of the
vector sector. For a non-trivial solution with $G_1 \ne 0$ we
calculate one-loop terms giving contribution to
equation for form-factor of vector terms. As a result  of the first
approximation we obtain just the isotopic vector terms, which are 
presented in expression~(\ref{intas}).

Let us note, that at this stage we have two possibilities: trivial 
solution $G_i = 0$ and non-trivial 
one~(\ref{Ci}, \ref{result1}, \ref{Cir}). 
We shall see below that a choice between the possibilities will 
be determined by QCD interaction. 

\section{ Scalar and pseudo-scalar states}

Now with the non-trivial solution of the compensation
equation we arrive at an effective theory in which
there are already no undesirable four-fermion terms in
{\bf free} Lagrangian~(\ref{addsub}) while they are
evidently present in {\bf interaction} Lagrangian~(\ref{intas}). 
Indeed four-fermion terms in these two
parts of the full Lagrangian differ in sign and
the existence of the non-trivial solution of
compensation equation for Lagrangian~(\ref{addsub})
means non-existence of the would be analogous equation,
formulated for signs of four-fermion terms in
{\bf interaction} Lagrangian~(\ref{intas}).\footnote{In other words 
the fact, that sum of a series
$\sum G^n a_n = 0$ for some value $G$, by no means leads to a conclusion,
that sum of the same series with $G \to -\,G$ vanishes as
well.}

So provided the non-trivial solution
is realized the compensated
terms go out from
Lagrangian~(\ref{addsub}) and we obtain the following
Lagrangian
\beq
& &L\,=\,\frac{\imath}{2}
\Bigl(\bar\psi\gamma_
\mu\partial_\mu\psi-\partial_\mu\bar\psi\gamma_\mu\psi
\biggr)\,-\,\frac{1}{4}\,F_{0\,\mu\nu}^aF_{0\,\mu\nu}^a\,-
\,m_0\bar\psi\,\psi\,+ \nn\\
& &+\,g_s\,\bar\psi\gamma_\mu
t^a A_\mu^a\psi\,-\,\frac{1}{4}\,\biggl( F_{\mu\nu}^aF_
{\mu\nu}^a -
F_{0\,\mu\nu}^aF_{0\,\mu\nu}^a\biggr)\,-\,\frac{G_1}{2}
\cdot\Bigl(\bar\psi \tau^b\gamma_5 \psi\,\bar\psi \tau^b
\gamma_5\psi -\bar\psi\,\psi\,\bar\psi\,\psi\biggr)\,-
\nn\\
& &-\,\frac{G_2}{2}\cdot\Bigl(\bar\psi \tau^b\gamma_\mu
\psi\,\bar\psi \tau^b\gamma_\mu \psi +
\bar\psi \tau^b\gamma_5 \gamma_\mu \psi \bar\psi \tau^b
\gamma_5 \gamma_\mu \psi\biggr)\,;\label{intas1}
\eeq
where $G_1$, $G_2$ are defined by relations~(\ref{result1}) and 
form-factor $F_1$ is defined by eqs.~(\ref{Ci}, \ref{Cir}).

Here we have to comment the meaning of the strong coupling constant
$g_s$. It is well-known that $g_s^2/4\,\pi = \alpha_s(q^2)$ is
the running constant depending on the momentum variable.
We need this constant in the low-momenta region. However the perturbation
theory in QCD does not work for small $q^2$. We assume that
in this region $\alpha_s(q^2)$ may be approximated by its
average value $\alpha_s$. This assumption is very close to
conception of frozen strong coupling at low momenta~\cite{sim}. The
consideration of average low-momenta value of $\alpha_s$~\cite
{sim, Shirk, dok, braz} leads to definition of a possible range of values
of $\alpha_s$ from $0.45$ up to $0.75$. So in what follows we use constant
$\alpha_s$ which is assumed to fit this interval of possible values.

Thus, bound state problems in the present approach are
formulated starting from Lagrangian (\ref{intas1}).

Let us write down Bethe-Salpeter
equation for a state in zero-spin (scalar and pseudo-scalar)
channel in the same approximation as was used in equation~(\ref{eq1}). 
Let us begin with massless states. The definitions of momenta are the 
same as in eq.(\ref{eq1})
\beq
& &\Psi(p^2)\,=\,\frac{G_1\,N}{2\,\pi^4}\,\int
\frac{ \Psi(q^2)
\,dq}{q^2}\,
+\,\frac{(G_1^2+6 G_1 G_2) N}{32\, \pi^6}\,\times\nn\\
& & \int\biggl(
2 \Lambda^2+(p-q)^2 \log\frac{(p-q)^2}{\Lambda^2}-\frac{ 3}{2} (p-q)^2\,-
\,\frac{m^4}{2\,(p-q)^2}
\biggr)\frac{ \Psi(q^2)\,dq}{q^2}\,.\label{eqpsi2}
\eeq
Here $m$ is a quark mass, which in general may not coincide with $m_0$.  
We define the value of $m$ after considering the spontaneous breaking 
of the chiral symmetry.

After angular integrations we obtain the one-dimensional 
equation similar to eq.(\ref{eq1xy})
\beq
& &\Psi(x) = \frac{G_1\,N}{2 \pi^2}\int_{m^2}^
{\infty} 
\Psi(y)\,dy +  \frac{(G_1^2+6 G_1 G_2) N}
{32\, \pi^4} \Biggl(\frac{1}{6\,x}\int_{m^2}^x 
(y^2 - 3 m^4) \Psi(y)\,
dy\,+\nn\\
& &+ \frac{3}{2}\int_{m^2}^x y \Psi(y)\,dy + 
\log\,x \int_{m^2}^x y \Psi(y)\,dy + x\,\log\,x \int_{m^2}^x 
\Psi(y)\,dy + 
\int_x^{\infty} y \log\,y\, \Psi(y)\,dy +\nn\\
& &+\,x \int_x^{\infty}  
\Bigl(\log\,y\,+\frac{3}{2}\biggr) \Psi(y)\,dy\,+\,
\frac{x^2 - 3 m^4}{6}
\int_x^{\infty}\,\frac{\Psi(y)}{y}\,dy + 
\Bigl(2\bar\Lambda^2-
\frac{3}{2}\, x\Bigr)\int_{m^2}^{\infty} \Psi(y)\,dy - \nn\\
& &\frac{3}{2}\,\int_{m^2}^{\infty} y \Psi(y)\,dy\,-\,\log\,
\Lambda^2 \Bigl(\int_{m^2}^{\infty} y \Psi(y)\,dy\,+\,
x \int_{m^2}^{\infty} \Psi(y)\,dy\Bigr)\,\Biggr)\,;
\label{eq1psi}
\nn
\eeq
The corresponding differential equation for $\Psi(x)$ is 
almost the same, as the previous one~(\ref{diff1}) with 
one essential difference. Namely the sign afore $\beta$ 
is opposite.
\beq
& &\Biggl(\biggl(z \frac{d}{dz}-\bar b\biggr)\biggl(z \frac
{d}{dz}-\bar a\biggr)\,z \frac{d}{dz}\,\biggl(z \frac{d}{dz}-
\frac{1}{2}\biggr)\biggl(z \frac{d}{dz}-\frac{1}{2}\biggr)
\biggl
(z \frac{d}{dz}-1\biggr) + \frac{\beta\,z}{2^6}\Biggr)
\,\Psi(z)\,=\,0\,;\nn\\
& &z\,=\,x^2\,;\quad \bar a\,=\,\frac{-1+\sqrt{1+64 u}}{4}\,;
\quad \bar b\,=\,\frac{-1-\sqrt{1+64 u}}{4}\,;\quad 
u\,=\frac{\beta m^4}{64}\,.\label{Meijerpsi}
\eeq
In this case we have the
following solution decreasing at
infinity
\beq
& &\Psi(z)\,=\,C^*_1\,G_{06}^{30}\Bigl(\,z\,
|1,\,\frac{1}{2},\,0,\,\frac{1}{2},\,\bar a,\,
\bar b\Bigr)\,+\,C_2^*\,G_{06}^{30}\Bigl(\,z\,|1,\,\frac{1}{2},\,\frac{1}
{2},\,0,\,\bar a,\,\bar b\Bigr)\,+\nn\\
& &+\,C_3^*\,G_{06}^{30}\Bigl(\,z\,|1,\,\bar a,\,
\bar b,\,\frac{1}{2},\,\frac{1}{2},\,0\Bigr)+\,C_4^*\,
G_{06}^{30}\Bigl(\,z\,|\frac{1}{2},\,\bar a,\,\bar b,\,1,\,\frac{1}{2},\,0
\Bigr)\,.\label{cpsi}\\
& &z\,=\,\frac{\beta\,x^2}{2^6}\,.\nn
\eeq
Constants $C_i^*$ are defined by the following conditions
\be
\int_{u}^{\infty}\frac{ \Psi(z)\,dz}{\sqrt{z}}\,=\,0\,;\quad
\int_{u}^{\infty} \Psi(z)\,dz\,=\,0\,;\quad
\int_{u}^{\infty} \sqrt{z}\,\Psi(z)\,dz\,=\,0\,;\quad
\Psi(u)\,=\,1.\label{boundpsi}
\ee
where $u= \beta\,m^4/2^6$. Let us remind, that boundary 
conditions~(\ref{boundpsi})
guarantee cancellation of terms in equation~(\ref{eqpsi2}) containing
cut-off $\Lambda$.
Performing integrations in expressions~(\ref
{boundpsi}), we have the following set of equations
\beq
& &C^*_1\,G_{06}^{30}\Bigl(\,u\,
|1,\,\frac{1}{2},\,0,\,\frac{1}{2},\,\bar a,\,
\bar b\Bigr)\,+\,C_2^*\,G_{06}^{30}\Bigl(\,u\,|1,\,\frac{1}{2},\,\frac{1}{2},
\,0,\,\bar a,\,\bar b\Bigr)\,+\nn\\
& &+\,C_3^*\,G_{06}^{30}\Bigl(\,u\,|1,\,\bar a,\,
\bar b,\,\frac{1}{2},\,\frac{1}{2},\,0\Bigr)+\,C_4^*\,
G_{06}^{30}\Bigl(\,u\,|\frac{1}{2},\,\bar a,\,\bar b,\,1,\,\frac{1}{2},\,0
\Bigr)\,=1\,;\nn\\
& &-\,C^*_1\,G_{06}^{30}\Bigl(\,u\,
|\frac{3}{2},\,1,\,\frac{1}{2},\,0,\,\frac{1}{2}+\bar a,\,
\frac{1}{2}+\bar b\Bigr)\,+\,C_2^*\Biggl(\frac{\Gamma(3/2)}{\Gamma(1/2)
\Gamma(1/2-\bar a)\Gamma(1/2-\bar b)}\,-\nn\\
& &-\,G_{17}^{31}\Bigl(\,u\,|^1_{3/2,\,1,\,1,\,1/2,
\,0,\,1/2+\bar a,\,1/2+\bar b}\Bigr)\Biggr)\,
-\,C_3^*\,G_{06}^{30}\Bigl(\,u\,|\frac{3}{2},\,\frac{1}{2}+\bar a,\,
\frac{1}{2}+\bar b,\,\frac{1}{2},\,1,\,0\Bigr)-\nn\\
& &-\,C_4^*\,
G_{06}^{30}\Bigl(\,u\,|1,\,\frac{1}{2}+\bar a,\,\frac{1}{2}+\bar b,\,\frac{3}{2},\,
\frac{1}{2},\,0\Bigr)\,=\,0\,;\nn\\
& &C^*_1\,\Biggl(\frac{\Gamma(3/2)}{\Gamma(-1/2)\Gamma
(-\bar a)\Gamma(-\bar b)}\,-\,G_{17}^{31}\Bigl(\,u\,
|^1_{2,\,3/2,\,1,\,3/2,\,0,\,1+\bar a,\,1+\bar b}\Bigr)\Biggr)\,-\nn\\
& &-\,C_2^*\,G_{06}^{30}\Bigl(\,u\,|2,\,\frac{3}{2},\,
\frac{3}{2},\,0,\,1+\bar a,\,1+\bar b\Bigr)\,
-\,C_3^*\,G_{06}^{30}\Bigl(\,u\,|2,\,1+\bar a,\,
1+\bar b,\,\frac{3}{2},\,\frac{3}{2},\,0\Bigr)\,-\nn\\
& &-\,C_4^*\,
G_{06}^{30}\Bigl(\,u\,|\frac{3}{2},\,1+\bar a,\,1+\bar b,\,2,\,
\frac{3}{2},\,0\Bigr)\,=\,0\,;\label{boundpsiu}\\
& &-\,C^*_1\,G_{17}^{31}\Bigl(\,u\,
|^1_{\frac{5}{2},\,2,\,\frac{3}{2},\,2,\,0,\,\frac{3}{2}+\bar a,\,
\frac{3}{2}+\bar b}\Bigr)\,+\,C_2^*\Biggl(\frac{\Gamma(5/2)}{\Gamma(-1/2)
\Gamma(-1/2-\bar a)\Gamma(-1/2-\bar b)}\,-\nn\\
& &-\,G_{17}^{31}\Bigl(\,u\,|^1_{5/2,\,2,\,
2,\,3/2,\,0,\,3/2+\bar a,\,3/2+\bar b}\Bigr)\Biggr)\,
-\,C_3^*\,G_{17}^{31}\Bigl(\,u\,|^1_{5/2,\,3/2+\bar a,\,
3/2+\bar b,\,2,\,2,\,3/2,\,0}\Bigr)\,-\nn\\
& &-\,C_4^*\,
G_{17}^{31}\Bigl(\,u\,|^1_{2,\,3/2+\bar a,\,3/2+\bar b,\,5/2,\,2,\,
3/2,\,0}\Bigr)\,=\,0\,.\nn
\eeq
For a given value of $u$ these conditions~(\ref{boundpsiu}) uniquely
define four coefficients $C_i^*$.  
The result, that equation~(\ref{eqpsi2}) has unique
solution, which satisfies all boundary conditions, corresponds to 
existence of a zero-mass state in the same approximation as is used  
for compensation equation~(\ref{eq1}). This is quite natural due to 
Bogolubov-Goldstone theorem~\cite{Bog2, Bog, Gold}.

However we have to take into account chromodynamic interaction as well as 
an interaction of these mesons ($\phi$ and $\pi_a$) with quarks. Indeed 
we have just shown the existence of this states and so the following 
effective meson-quark interaction is to exist
\be
-\,g\,\Bigl(\phi\,\bar\psi\,\psi+\imath\,\pi_a\,\bar\psi\,
\gamma_5\,\tau_a\psi\Bigr)\,;\label{sigpsi}
\ee
where $g$ is defined by normalization condition of zero-spin
states
\be
\frac{g^2 \,N}{4\,\pi^2}\,I_2\,=\,1\,; \quad I_2\,=\,
\int_{m^2}^\infty
\frac{\Psi(p^2)^2\,dp^2}{p^2}\,=\,\int_{u}^\infty
\frac{\Psi(z)^2\,dz}{2\,z}\,.\label{g}
\ee
The form-factor of interaction~(\ref{sigpsi}) for our standard quark 
momenta prescription ($p,\,-\,p$) is Bethe-Salpeter wave function defined 
by eqs.(\ref{cpsi}, \ref{boundpsiu}). 
The account of contributions of meson-quark interaction was considered
in the framework of the Nambu -- Jona-Lasinio model {\it e.g.} 
in works~\cite{ebvol, Anph} and was shown to be corresponding to the 
next order of the $1/N$ expansion.
Let us calculate a mass correction term due to these contributions.
For the purpose let us take into account terms of the
first order in $P^2$, where $P$ is the momentum of
a scalar (and pseudo-scalar) meson and one-loop
terms being due to quark-gluon QCD interaction and quark-meson vertices. 
Note that for the last loops we use massless meson exchange.  
We define momenta of left-hand legs in fig. 2 to be $p + P/2$ and 
$-p + P/2$ and obtain the following equation  
\beq
& &\Psi_P(p^2)\,=\,\frac{G_1\,N}{2\,\pi^4}\,\int
\frac{ \Psi_P(q^2)
\,dq}{q^2}\biggl(1 - \frac{3\,P^2}{4\,q^2} +
\frac{(q P)^2}
{(q^2)^2} \biggr)\,
+\,\frac{(G_1^2+6 G_1 G_2) N}{32\, \pi^6}\,\times\nn\\
& & \int\biggl(
2 \Lambda^2+(p-q)^2 \log\frac{(p-q)^2}{\Lambda^2}-\frac{ 3}{2} (p-q)^2\,
-\,\frac{\mu^2}{2\,(p-q)^2}
\biggr)\biggl(1 - \frac{3\,P^2}{4\,q^2} + \frac{(q P)^2}
{(q^2)^2} \biggr)\times\nn\\
& &\times\frac{ \Psi_P(q^2)\,dq}{q^2}\,
+\,\Bigl(\frac{g_s^2}{4\,\pi^4}+\frac{g^2}{8\,\pi^4}\Bigr)\,\int
\frac{ \Psi_P(q^2)\,dq}
{q^2 (q-p)^2}\,.\label{eqpsi1}
\eeq

In the course of QCD term calculation we use
transverse Landau gauge \footnotemark[1]
\footnotetext[1]{In the approximation used the transverse
gauge leads to absence of renormalization of both
vertex and spinor field.}. Let us multiply equation~(\ref
{eqpsi1}) by $\Psi_P(p^2)/p^2$ at $P = 0$ and integrate by
$ p $. Due to equation~(\ref{eqpsi2}) be satisfied we
have
\beq
& &-\,\frac{P^2}{2}\,\int \frac{ \Psi(q^2)^2\,dq}{(q^2)^2}
\,+\,\Bigl(\frac{g_s^2}{4\,\pi^4}\,+\,\frac{g^2}{8\,\pi^4}\Bigr)\,\int
\frac{ \Psi(p^2)\,dp}
{p^2} \int \frac{ \Psi(q^2)\,dq}{q^2 (q-p)^2}\,=\,0\,;
\label{P2}
\eeq
After angular integration we get
\beq
& &\frac{P^2\,\pi^2}{2}\,I_2\,=\,\frac{2\,g_s^2+g^2 }{8}\,
\int_{m^2}^\infty\,\Psi(x)\,dx\,\biggl(\frac{1}{x}\int_
{m^2}^x\,\Psi(y)\,dy\,+\,\int_x^\infty
\frac{\Psi(y)\,dy}{y}\,\biggr)\,=\nn\\
& &=\,\frac{2 g_s^2+ g^2}{2\,\sqrt{\beta}}\,
\int_u^\infty\,\,\frac{\Psi(z)\,dz}{z}\int_{u}
^z\,\frac{\Psi(t)\,dt}{\sqrt{t}}\,=\,\frac{(2 g_s^2+g^2)\,I_5}
{2\,\sqrt{\beta}};
\; z\,=\frac{\beta\,x^2}{64}\,;\;
t\,=\frac{\beta\,y^2}{64}\,.\label{eigen}
\eeq
Integral inside of $I_5$ with account of boundary conditions~(\ref
{boundpsi}) reads
\beq
& &\int_u^z \frac{\Psi(t)\,dt}{\sqrt{t}}\,=\,C_1^*\,
G_{06}^{30}\Bigl(
z\,|\,\frac{1}{2},\,1,\,\frac{3}{2},\,0,\,\frac{1}{2}+\bar a,\,
\frac{1}{2}+\bar b\Bigr)\,-\nn\\
& &-\,C_2^*\,G_{06}^{30}\Bigl(z\,|\,0,\,1,\,
\frac
{3}{2},\,\frac{1}{2},\,\frac{1}{2}+\bar a,\,\frac{1}{2}+\bar b\Bigr) +
C_3^*\,G_{06}^{30}\Bigl(z\,|\,\frac{3}{2},\,\frac{1}{2}+\bar a,\,
\frac{1}{2}+\bar b,\,0,\,\frac{1}{2},\,1\Bigr)+\nn\\
& &+\,C_4^*\,G_{06}^{30}\Bigl(z\,|\,1,\,\frac{1}{2}+\bar a,\,\frac{1}{2}+
\bar b,\,0,\,\frac{1}{2},\,\frac{3}{2}\Bigr)\,;\label{cpsiint}
\eeq
and after substitution of relation~(\ref{cpsiint}) into integral $I_5$ it
is to be calculated numerically. Note that while evaluating
integral~(\ref{cpsiint}) we use the following relation, which is
presented in work~\cite{Arb05}
\be
G_{17}^{31}\Bigl(z\,|^1_{1,\,c,\,d,\,0,\,g,\,a,\,b}
\Bigr)\,=\,\frac{\Gamma(c) \Gamma(d)}{\Gamma(1-g)
\Gamma(1-a)\Gamma(1-b)}\,-\,G_{06}^{30}\Bigl(z\,|0,\,c,\,
d,\,g,\,a,\,b\Bigr)\,.\label{relation}
\ee

Integral $I_5$ turns to be positive, so the mass squared of scalar and
pseudo-scalar mesons is shifted to negative value
\be
m_t^2\,=\,-\,\biggl(\frac{\alpha_s}{\pi}\,+\,\frac{g^2}{8\,\pi^2}
\biggr) \frac{8\,I_5}{\sqrt{\beta}\,I_2}\,.\label{mt}
\ee

\section{Spontaneous breaking of the chiral symmetry}

The negative value of $m_t^2$~(\ref{mt}) means instability of
the vacuum. Therefore we have to consider an effective potential
depending on scalar field $\phi$. In doing this we need an expression
for mass operator of the quark $\Sigma(p^2)$. The Schwinger-Dyson 
equation defining this function in our approximation reads as follows
\beq
& &\Sigma(p^2)\,=\,m_0\,+\frac{G_1\,N}{2\,\pi^4}\,\int
\frac{ \Sigma(q^2)
\,dq}{q^2}\,
+\,\frac{(G_1^2+6 G_1 G_2) N}{32\, \pi^6}\,\times\nn\\
& & \times\,\int\biggl(
2 \Lambda^2+(p-q)^2 \log\frac{(p-q)^2}{\Lambda^2}-\frac{ 3}{2} (p-q)^2\,
-\,\frac{m^4}{2\,(p-q)^2}
\biggr)\,\frac{ \Sigma(q^2)\,dq}{q^2}\,+\nn\\
& &+\,\Bigl(\frac{g_s^2}{4\,\pi^4}+\frac{g^2}{8\,\pi^4}\Bigr)\,\int
\frac{ \Sigma(q^2)\,dq}
{q^2 (q-p)^2}\,.\label{eqsig1}
\eeq
The first approximation corresponds to $m_0 = g_s = g =0$. 
Then equation~(\ref{eqsig1}) exactly coincides with equation~(\ref{eqpsi2})
for Bethe-Salpeter wave function $\Psi(p^2)$~(\ref{cpsi}). Similar 
situation takes place in standard NJL model~\cite{Vol2}. For
non-zero $m_0$ we have without gluon and meson corrections
\be
\Sigma(x)\,=\,m_0\,+\,(m-m_0)\,\Psi(x)\,;\quad
\Sigma(-m^2)\,=\,m\,.\label{sigm}
\ee
Emphasize that approximate solution~(\ref{sigm}) of
equation~(\ref{eqsig1})
exists for any value of $m$. For definition of $m$ one has to 
turn to the spontaneous breaking of the chiral symmetry. 

Let us write down the effective potential which defines a possibility 
of the symmetry breaking. We look for terms proportional
to $\phi^n$ for $n\,=\,1,\,2,\,3,\,4$. The term with $n\,=\,2$ is
evidently defined by~(\ref{mt}). For terms with $n\,=\,3,\,4$ we
take quark-loop diagrams with three and four scalar legs respectfully
and as a result we have the following effective potential
\beq
& &V\,=\,m^4\,\Biggl(-\,\Bigl(1-\Bigl(\frac{u_0}{u}\Bigr)^{1/4}\Bigr)\,
\Bigl(\frac{1}{8\,\pi^2}+\frac{\alpha_s}{\pi\,g^2}\Bigr)\,\frac{I_5\,\xi}
{\sqrt{u}\,I_2}\,-\,\Bigl(\frac{1}{8\,\pi^2}+\frac{\alpha_s}
{\pi\,g^2}\Bigr)\,\frac{I_5\,\xi^2}{2\,\sqrt{u}\,I_2}\,+\nn\\
& &+\,\frac{3\,\xi^3}
{2\,\pi^2}\,\Bigl(\Bigl(\frac{u_0}{u}\Bigr)^{1/4}\,I_3\,+\,\Bigl(1-
\Bigl(\frac{u_0}{u}\Bigr)^{1/4}\Bigr)\,I_4\Bigr)\,+\,\frac{3\,\xi^4}{8\,
\pi^2}\,I_4\Biggr)\,;\quad \xi\,=\,\frac{g\,\phi}{m}\,.\label{V}
\eeq
Here
\be
I_3\,=\,\int_u^\infty\,\frac{\Psi(z)^3\,dz}{2\,z}\,;\qquad
I_4\,=\,\int_u^\infty\,\frac{\Psi(z)^4\,dz}{2\,z}\,.\label{I34}
\ee
The connection between terms with $n = 1$ and $n = 2$
is obtained from the fact, that the tadpole term due to eq.(\ref{sigm})
gives just the same
contribution as the two-loop one up to factor $\,(m-m_0)/g$. The
contribution to the tadpole term being proportional to $m_0$ is zero
due to boundary conditions~(\ref{boundpsi}).

As for one-loop terms with $n \geq 5$, they all converge with
point-like vertices.
In this case they can be calculated and summed up to give the following
additional term
\be
\Delta\,V\,=\,m^4\,\biggl(\frac{1}{16\,\pi^2}\,(1-\xi)^4\,\log|1-\xi|\,+\,
\frac{\xi}{16\,\pi^2}\,-\,\frac{7\,\xi^2}{32\,\pi^2}\,+\,\frac{13\,\xi^3}
{48\,\pi^2}\,-\,\frac{25\,\xi^4}{192\,\pi^2}\biggr)\,;\label{V5}
\ee
which evidently does not destroy stability conditions and turns to
influence results quite insignificantly. 
Thus we neglect it.

We look for a minimum of potential~(\ref{V}) that is for a solution of
the following equation
\be
\frac{\partial\,V}{\partial\,\xi}\,=\,0\,.\label{dV}
\ee
Constituent quark mass is expressed through the vacuum expectation value 
of scalar field $\phi$
\be
m\,=\,m_0\,+\,g\,\eta\,;\quad \eta\,=\,<\,\phi\,>\,.\label{meta}
\ee
Bearing in mind definitions~(\ref{A0}, \ref{cpsi}) of parameters
$u_0$ and $u$,
we come to the conclusion, that the position of minimum $\xi_0$ has
to be the following
\be
\xi_0\,=\,\Biggl(1-\Bigl(\frac{u_0}{u}\Bigr)^{1/4}\Biggr)\,.
\label{xi0}
\ee
Thus from relations~(\ref{g}, \ref{V}, \ref{dV}, \ref{xi0}) we obtain
the following expression for $\alpha_s$
\be
\alpha_s\,=\,\frac{\pi\,\sqrt{u}}{I_5}\biggl(1 - \biggl(\frac{u_0}{u}\biggr)^
{\frac{1}{4}}\biggr)\,\Biggl(3\,\biggl(\frac{u_0}{u}\biggr)^
{\frac{1}{4}}\,I_3\,+\,4\,\biggl(1 - \biggl(\frac{u_0}{u}\biggr)^
{\frac{1}{4}}\biggr)\,I_4\Biggr)\,-\,\frac{\pi}{6\,I_2}\,.\label{alu}
\ee
Here all integrals are functions of $u$ and so relation~(\ref{alu})
defines  function $\alpha_s(u)$.

Now it is the proper place to comment the problem of stability.
From the very beginning we have two solutions: the trivial one
$G_1 = G_2 = 0,\,m = m_0$ and the non-trivial one, which in details
is presented above. We get convinced that the non-trivial solution
corresponds to the minimal negative value of effective
potential~(\ref{V}) while the trivial solution corresponds to its
value zero. So we are to conclude, that just the non-trivial
solution is stable and thus the non-trivial solution is to describe
the observable physical quantities.

We apply quark mass operator~(\ref{sigm}) to obtain  also
the expression for pion decay constant $f_\pi$. Considering one-loop
quark diagram for decay amplitude of process
$\pi^+ \to \mu^+\,\nu_\mu$ we have
\beq
& &f_\pi\,=\,\frac{g\,N}{4\,\pi^2}\int_{m^2}^\infty
\Bigl((m-m_0)\,\Psi(y)^2\,+\,m_0\,\Psi(y)\Bigr)
\frac{dy}{y}\,=\nn\\
& &=\,\frac{g\,N}{4\,\pi^2}\Bigl((m-m_0)\,I_2
+\,m_0\,I_1\Bigr)\,;\quad I_1\,=\,\int_u^\infty\frac
{\Psi(z)\,dz}{2\,z}\,.\label{fpi}
\eeq
Provided either $m_0 = 0$ or $I_2 = I_1$ we get
with account of normalization condition~(\ref{g}) just
the original Goldberger -- Treiman relation
$m = g\,f_\pi$. We use full relation~(\ref{fpi}). However
let us note, that values of the two integrals  are close
$I_2 \simeq I_1$ and the simple original relation works
with sufficient accuracy\footnote{Additional contributions to $f_\pi$ 
being due to the P-exponent, which maintain the electroweak gauge 
invariance, are shown to cancel in the chiral limit~\cite{ORV}.}.

\section{Pion mass and quark condensate}

In relations~(\ref{V}, \ref{fpi}) we have used approximation~(\ref{sigm})
for the quark mass operator.
For calculation of the pion mass and the quark condensate we
need the next approximation for mass operator. In view of this
we reformulate equation~(\ref{eqsig1}) for the following
function
\be
\Phi(p^2)\,=\,\frac{\Sigma(p^2)-m_0}{m-m_0}\,;\label{Phi}
\ee
and the first approximation for $\Phi$ is just $\Psi$.
Then we introduce~(\ref{Phi}) into~(\ref{eqsig1}) to obtain
\beq
& &\Phi(p^2)\,=\,\frac{G_1\,N}{2\,\pi^4}\,\int
\frac{((m-m_0)\,\Phi(q^2)\,+\,m_0)
\,dq}{(m-m_0)\,q^2}\,
+\,\frac{(G_1^2+6 G_1 G_2) N}{32\, \pi^6}\,\times\nn\\
& & \int\biggl(
2 \Lambda^2+(p-q)^2 \log\frac{(p-q)^2}{\Lambda^2}-\frac{ 3}{2} (p-q)^2\,
-\,\frac{m^4}{2\,(p-q)^2}
\biggr)\,\frac{((m-m_0)\,\Phi(q^2)\,+\,m_0)\,dq}{(m-m_0)\,q^2}\,+\nn\\
& &+\,\Bigl(\frac{g_s^2}{4\,\pi^4}+\frac{g^2}{8\,\pi^4}\Bigr)\,\int
\frac{(m_0\,+\,(m-m_0)\,\Phi(q^2))\,dq}
{(m-m_0)\,q^2 (q-p)^2}\,.\label{eqphi1}
\eeq
Now we subtract equation~(\ref{eqpsi1}) from equation~(\ref{eqphi1}) and
obtain the following relation
\beq
& & D(p^2)\,=\,\frac{G_1\,N}{2\,\pi^4}\,\Biggl(\int
\frac{((m-m_0)\,D(q^2)\,+\,m_0)
\,dq}{(m-m_0)\,q^2}\,+\,\int\frac{\Psi(q^2)\,dq}{q^2}\biggl(\frac{3\,P^2}
{4\,q^2} - \frac{(qP)^2}{(q^2)^2}\biggr)\Biggr)\,+\nn\\
& &+\,\frac{(G_1^2+6 G_1 G_2) N}{32\, \pi^6}\,\Biggl(
 \int\biggl(
2 \Lambda^2+(p-q)^2 \log\frac{(p-q)^2}{\Lambda^2}-\frac{ 3}{2} (p-q)^2\,
-\,\frac{m^4}{2\,(p-q)^2}
\biggr)\,\times\nn\\
& &\times\,\frac{((m-m_0) D(q^2)+m_0)\,dq}{(m-m_0)\,q^2}\,+
\,\int\frac{\Psi(q^2)\,dq}{q^2}\,\biggl(\frac{3\,P^2}
{4\,q^2} - \frac{(qP)^2}{(q^2)^2}\biggr)\times\nn\\
& &\times\,\biggl(2 \Lambda^2+(p-q)^2 \log\frac{(p-q)^2}{\Lambda^2}-
\frac{ 3}{2} (p-q)^2\,-\,\frac{m^4}{2\,(p-q)^2}
\biggr)\,\Biggr)+\,\nn\\
& &+\,\Bigl(\frac{g_s^2}{4\,\pi^4}+\frac{g^2}{8\,\pi^4}\Bigr)\,\int
\frac{((m-m_0)\,D(q^2)+m_0)\,dq}
{(m-m_0)\,q^2 (q-p)^2}\,.\label{eqphi2}\\
& & D(p^2)\,=\,\Phi(p^2)\,-\Psi(p^2)\,.\nn
\eeq
The analogous procedure is applied in standard NJL model~\cite{Vol2}, 
\cite{ERV} while 
proving the Gell-Mann, Oaks and Renner theorem~\cite{GM}.    
Then we again multiply~(\ref{eqphi2}) by $\Psi(p^2)/p^2$ and integrate
over $dp$. Due to equation~(\ref{eqpsi1}) be satisfied only
terms being proportional either to $P^2 = -\,m_\pi^2$ or to $m_0$ do not
cancel and finally we have the
following relation for mass of the $\pi$-meson 
\beq
& &m_\pi^2\,=\,\frac{m^2\,m_0}{2\,\pi\,(m-m_0)\,I_2\,\sqrt{u}\,}\biggl(
\alpha_s\,+\,\frac{g^2}{8\,\pi}\biggr)\,I_{\log}\,;\label{mpi}\\
& &I_{\log}\,=\,-\,\int_u^\infty\frac{\log\,z}{\sqrt{z}}\,\Psi(z)\,dz\,=
\,C_1^*\,G^{30}_{06}
\bigl(\,u\,|\frac{3}{2},\,\frac{1}{2}, 0,\,\frac{1}{2}+a,\,\frac{1}{2}+b,
\,0\bigr)\,-\nn\\
& &-\,C_2^*\,G^{30}_{06}\bigl(\,u\,|\frac{3}{2},\,0,
\,0,\, \frac{1}{2},\,\frac{1}{2}+a,\,\frac{1}{2}+b\bigr)\,-
\,C_3^*\,G^{30}_{06}\bigl(\,u\,|\frac{3}{2},\,\frac{1}{2}+a,\,\frac{1}{2}+
b,\,\frac{1}{2},\,0,\,0\bigr)\,+\nn\\
& & +\,C_4^*\,G^{30}_{06}\bigl(\,u\,|0,\,\frac{1}{2}+a, \frac{1}{2}+b,\,
\frac{3}{2},\,\frac{1}{2} ,\,0\bigr)\,.\nn
\eeq
We see that pion mass squared is proportional to $m_0$ in accordance to
the result of well-known work~\cite{GM}.
Note, that contributions being proportional to $m_0$ arising from
the first two terms of equation~(\ref{eqphi2}) are summed to overall
zero due to consequences of boundary conditions~(\ref{boundpsiu}).

From equation~(\ref{Phi}) we obtain also the next approximation
for $\Phi$, which leads to a non-zero value of the quark
condensate
\beq
& &<\bar q\,q>\,=\,-\,\frac{4\,N}{(2\,\pi)^4}\,\int\,\frac
{\Sigma(q)-m_0}{q^2 + m^2}\,dq\,=\nn\\
& &=\,\frac{N\,(m-m_0)}{\pi^2\,\beta}\,\biggl(
\frac{\alpha_s}{\pi}+
\frac{g^2}{8\,\pi^2}\biggr)\,\int_u^\infty\frac{dz}{\sqrt{z}}\,
\biggl(\int_u^z \frac{\Psi(t)\,dt}{\sqrt{t}}+\int_z^\infty \frac{\Psi(t)
\,dt}
{t}\biggr)\,=\label{bqq}\\
& &=\,-\,\frac{N\,(m-m_0)}{\pi^2\,\beta}\,\biggl( \frac{\alpha_s}{\pi}+
\frac{g^2}{8\,\pi^2}\biggr)\,\biggl(\int_u^z\frac{\Psi(t)\,\log t\,dt}
{\sqrt{t}}+\,2\,\sqrt{u}\int_u^\infty\frac{\Psi(t)\,dt}
{t}\biggr)\,.\nn
\eeq
After evaluating the integrals we have
\beq
& &<\bar q\,q>\,=\,-\,\biggl(\alpha_s+\frac{g^2}{8\,\pi}\biggr) \frac
{3\,m^2\,(m-m_0)}{8\,\pi^3\,\sqrt{u}}\,\biggl(-C_1^*\,\Bigl(G^{30}_{06}
\bigl(\,
u\,|\frac{3}{2},\,\frac{1}{2}, 0,\,0,\,\frac{1}{2}+a,\,\frac{1}{2}+b\bigr)
\,-\nn\\
& &-\,2\,G^{30}_{06}\bigl(\,u\,|1,\,\frac{1}{2}, \frac{1}{2},\,0,\,\frac
{1}{2}+a,\,\frac{1}{2}+b\bigr)\Bigr)\,+\nn\\
& &+\,C_2^*\,\Bigl(G^{30}_{06}\bigl(\,u\,|\frac{3}{2}, 0, 0, \frac{1}{2},
\frac{1}{2}+a,\,\frac{1}{2}+b\bigr)+2\,G^{30}_{06}\bigl(\,u\,|1,\,1,
\frac{1}{2},\,\frac{1}{2},\,\frac{1}{2}+a,\,\frac{1}{2}+b\bigr)\Bigr)\,
+\label{qq}\\
& &+\,C_3^*\,\Bigl(G^{30}_{06}\bigl(\,u\,|\frac{3}{2},\,\frac{1}{2}+a,
\frac{1}{2}+b,\,\frac{1}{2},\,0,\,0\bigr)+2\,G^{30}_{06}\bigl(\,u\,|
\frac{1}{2},\,\frac{1}{2}+a, \frac{1}{2}+b,\,1,\,1,\,\frac{1}{2}\bigr)
\Bigr)\,-\nn\\
& &-\,C_4^*\,\Bigl(G^{30}_{06}\bigl(\,u\,|\frac{1}{2}+a,\,\frac{1}{2}+b, 0,
\,\frac{3}{2},\,\frac{1}{2},\,0)+2\,G^{30}_{06}\bigl(\,u\,|1,\,\frac{1}{2}
+a,\, \frac{1}{2}+b,\,1,\,\frac{1}{2},\,\frac{1}{2}\bigr)\Bigr)\biggr)\,;
\nn
\eeq

Scalar field $\phi$ corresponds to the $\sigma$-meson.
To calculate mass of the $\sigma$-meson we use relation for difference
of $\sigma$ and $\pi$ masses squared following from one-loop diagram
\be
m_\sigma^2\,-\,m_\pi^2\,=\,\frac{g^2\,N}{\pi^4}\int\,\frac
{\Sigma(q)^2\,\Psi(q)^2}{(q^2+m^2)^2}\,dq\,.\label{sig-pi}
\ee
Following again our rules and using expression~(\ref{sigm}) we have for
mass of the $\sigma$-meson
\be
m_\sigma^2\,=\,m_\pi^2\,+\,\frac{N\,g^2}{\pi^2}\,\biggl(m^2_0\,I_2\,+\,
2\,m_0(m-m_0)\,I_3\,+(m-m_0)^2\,I_4\biggr)\,;\label{msigma}
\ee
The $\sigma\,\pi\,\pi$ vertex gives according to triangle one-loop 
diagram the following coupling constant

\be
g_{\sigma \pi \pi}\,=\,\frac{g^3\,N}{\pi^2}\,\biggl(m_0\,I_3\,+\,(m-m_0)\,I_4\biggr)\,;\label{gsp}
\ee
and the $\sigma$-meson width reads
\be
\Gamma_\sigma\,=\frac{3\,g_{\sigma \pi \pi}^2}{16\,\pi\,m_\sigma^2}\,
\sqrt{m_\sigma^2-4\,m_\pi^2}\,.\label{gamsig}
\ee

\section{Numerical results and discussion}

Now we have expressions for all quantities under study.
Then we proceed as follows.\\
1) We calculate function $\alpha_s$~(\ref{alu}) depending on parameter
$u$~(\ref{cpsi}) and
get convinced, that the interesting range of $\alpha_s$ corresponds
to $u$ varying in the following region
\be
0.0005\,<\,u\,<\,0.0015\,.\label{u}
\ee
In doing this we use parameter $u_0 = 2\,10^{-8}$ according to
relation~(\ref{result1}) and  calculate constants $C^*_i,\,i=1,2,3,4\,$
from boundary conditions~(\ref{boundpsiu}) thus defining $\Psi(z)$.
Having $\Psi(z)$ we calculate integrals $I_j,\,j=1,2,3,4,5$.\\
2) We fix value $f_\pi= 93\,MeV$.\\
3) Then for given $u$ in range~(\ref{u}) from~(\ref{fpi}) we obtain
constituent quark mass $m$.\\
4) Having $m$ and $\alpha_s$ we calculate $m_\pi$ from~(\ref{mpi}).

For $u$ in range~(\ref{u})  $m_\pi$ varies insignificantly between $134\, MeV$
and $135\, MeV$ with maximal value $134.8\,MeV$ at $u\,=\,0.0009$, that
corresponds to $\alpha_s = 0.673$ and $m_0 = 20.27\,MeV$.
Considering this value of $m_\pi$ to be the most suitable, we present a
set of calculated parameters for this conditions including
quark condensate~(\ref{qq}) and parameters of the $\sigma$-meson~
(\ref{msigma}), (\ref{gamsig}) as well
\beq
& &\alpha_s\,=\,0.673\,;\qquad m_0\,=\,20.3\,MeV\,; \nn\\
& &m_\pi\,=\,135\,MeV\,;\quad f_\pi\,=\,93\,
MeV\,;\quad m_\sigma\,=\,492\,MeV\,;
\quad \Gamma_\sigma\,=\,574\,MeV\,\nn\\
& &m\,=\,295\,MeV\,;\qquad <\bar q\,q>\,=\,-\,(222\,MeV)^3\,;\label{res}\\
& &G_1\,=\,\frac{1}{(244\,MeV)^2}\,;\qquad g\,=\,3.16\,.\nn
\eeq
The upper line here is our input, while all other quantities are
calculated from these two fundamental parameters. We present here also 
values of four-fermion constant $G_1$ and of meson-quark coupling 
$g$. Note, that for these  
calculations uncertainties due to our method of infrared cut-off are 
defined not by eq.(\ref{acm0}) but by the following quantity
$$
\sqrt{u}\,=\,\sqrt{0.0009}\,=\,3\,10^{-2}\,;
$$ 
that is the accuracy of numbers~(\ref{res}) is evidently not better 
than 3\%. There are also other sources of uncertainties and so 
we may estimate the overall accuracy to be of order of 10\%. The main 
contribution to this estimate is provided by the next orders of $1/N$ 
expansion, according to the discussion in the Section 2.

Bearing in mind the last remarks, we may consider the correspondence of 
our results for $m_\pi$, $f_\pi$ and $<\bar q\,q>$
to existing data being quite satisfactory. Value for constituent quark 
mass is
also consistent. As for parameters of the $\sigma$-meson,
experimental data according to~\cite{pdg} give a wide range  for their
possible values. However let us
note, that recent determinations~\cite{lewt, bugg} of the
$\sigma$-meson parameters give more definite results. They are
respectfully the following
\be
m_\sigma\,=\,470 \pm 30\,,\quad \Gamma_\sigma\,=\,590 \pm 40\,;\qquad
m_\sigma\,=\,541 \pm 39\,,\quad \Gamma_\sigma\,=\,504 \pm 80\,;\label{sex}
\ee
that agrees with the present calculations. There is a recent
analysis of the $\pi-\pi$ data with light $\sigma$~\cite{anis},
which also agrees with $\sigma$ parameters~(\ref{res}).

To conclude we would like to emphasize that the present approach
for the first time permits to determine parameters of
effective interaction inherent to the Nambu -- Jona-Lasinio model in
terms of parameters of the fundamental QCD. The optimal value of
$\alpha_s\,=\,0.67$ in~(\ref{res}) is quite reasonable from the point of
view of the existing knowledge on its low-momenta behaviour
(see again~\cite{Shirk}). As for value of current quark mass
$m_0\simeq 20\,MeV$, it seems to be rather larger than usual
values $m_0(2\, GeV)\simeq 4-8\,MeV$. To comment the situation let us
note, that firstly the low value of $m_0$ being mentioned corresponds to
perturbative region and the problem how this running parameter varies
while $\sqrt{q^2}$ moves to low energy region deserves a special study. 
In considering of the running $m_0$
we have to take into account the effective NJL interaction as well.
Secondly, the lattice studies
give as a rule rather high values for $m_0$, {\it e.g.} in work~\cite
{lattice} values of $m_0$ corresponds just to few tens of $MeV$. The
smaller values are to be obtained in the continuous
limit, which till now is performed only by an extrapolation procedure. 

So we may state that the aim of the work
is achieved. We have begun with the demonstration of
the non-trivial solution of the compensation equation.
The appearance of scalar and pseudo-scalar excitations
(mesons) in the same approximation is a consequence of
its existence. The account of QCD interaction and of meson-quark
interaction leads to
the shift of their masses squared to the negative
region, i.e. to the appearance of tachyons, which are
necessary for scalar condensate to arise. As a result
we obtain the standard scheme leading to the spontaneous chiral
symmetry breaking.
Subsequent approximations of the approach are related to values
of the quark condensate and of the pion mass.

We have shown that the application of the method of
works~\cite{Arb05, Arb04}, which is based on Bogolubov
 approach, to the low-energy region of
hadron physics leads to quite reasonable results. Let us
once more emphasize that we have no additional parameters but those
entering in the low-energy QCD: $\alpha_s$ and $m_0$. Thus
we derive effective interaction of NJL type from the fundamental QCD. 
Basing on the results
of the work  we would make two essential conclusions.

Firstly, a subsequent development of the present approach
to the hadron physics quite deserves attention. In
particular it is advisable to apply the approach to
calculation of parameters of vector mesons
$\rho,\,\omega,
\,A_1$, to consider hadrons containing $s$-quark, to
take into account the $\pi-A_1$-mixing, to study
diquarks {\it etc.}. These problems comprise subjects
for forthcoming studies.

Secondly, the positive result of applicability test with
Nambu -- Jona-Lasinio model, being taken as an example,
allows to hope for successful application of the approach
to other problems. In particular we mean the problem
of a dynamical breaking of the electroweak symmetry.
A qualitative discussion of possible variants in this
direction is presented {\it e.g.} in work~\cite{SUSY}.

\section{Acknowledgments}

The authors express their gratitude to V.I. Savrin and D.V. Shirkov
for valuable discussions. The work was supported in part by
RFBR grant 05-02-16699.

\newpage
\begin{center}
{\bf Figure captions}
\end{center}
\bigskip
\bigskip
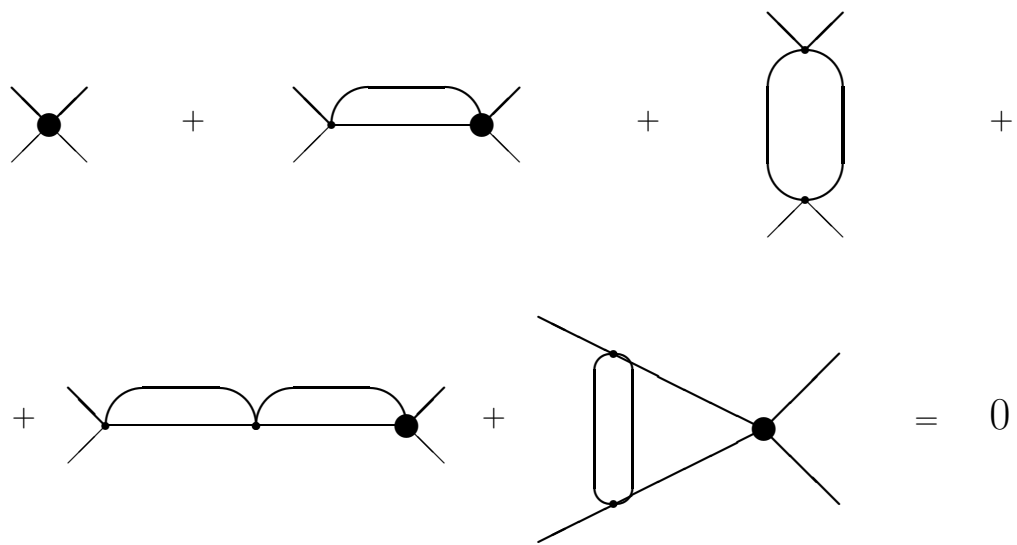
Fig. 1. Diagram representation of the compensation
equation. Black spot corresponds to four-fermion
vertex with a form-factor. Simple point corresponds to
a point-like vertex.\\
\\
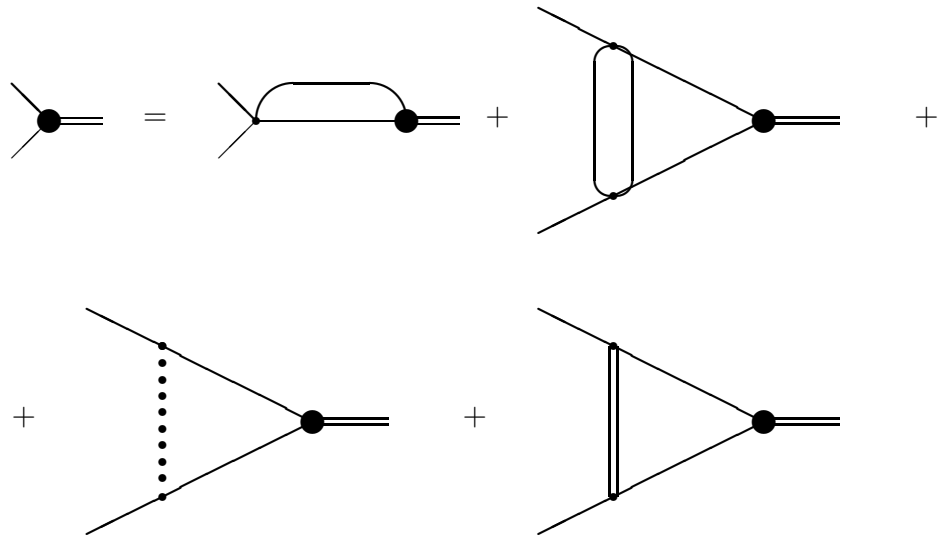
Fig. 2.
Diagram representation of Bethe-Salpeter equation
for scalar and pseudo-scalar bound state. A meson corresponds to a
double line. Dotted line corresponds to the gluon.

\newpage
\begin{picture}(160,65)
{\thicklines
\put(5,50.5){\line(-1,1){5}}
\put(5,50.5){\line(1,1){5}}
\put(5,50.5){\circle*{3}}}
\put(5,50.5){\line(-1,-1){5}}
\put(5,50.5){\line(1,-1){5}}

\put(22.5,50){+}
{\thicklines
\put(42.5,50.5){\line(-1,1){5}} \put(52.5,50.5)
{\oval(20,10)[t]}
\put(62.5,50.5){\line(1,1){5}}\put(42.5,50.5)
{\circle*{1}}
\put(62.5,50.5){\circle*{3}}}
\put(42.5,50.5){\line(-1,-1){5}} \put(62.5,50.5)
{\line(1,-1){5}}
\put(42.5,50.5){\line(1,0){20}}
\put(83,50){+}

{\thicklines
\put(105.5,60.5){\line(-1,1){5}}
\put(105.5,60.5){\line(1,1){5}}
\put(105.5,50.5){\oval(10,20)}
\put(105.5,60.5){\circle*{1}}
\put(105.5,40.5){\circle*{1}}}
\put(105.5,40.5){\line(-1,-1){5}}
\put(105.5,40.5){\line(1,-1){5}}
\put(130,50){+}
\put(0,10.5){+}
{\thicklines
\put(12.5,10.5){\line(-1,1){5}} \put(22.5,10.5)
{\oval(20,10)[t]}
\put(12.5,10.5)
{\circle*{1}}
\put(32.5,10.5){\circle*{1}}}
\put(12.5,10.5){\line(-1,-1){5}}
\put(12.5,10.5){\line(1,0){20}}
{\thicklines
 \put(42.5,10.5)
{\oval(20,10)[t]}
\put(52.5,10.5){\line(1,1){5}}\put(32.5,10.5)
{\circle*{1}}
\put(52.5,10.5){\circle*{3}}}
 \put(52.5,10.5)
{\line(1,-1){5}}
\put(32.5,10.5){\line(1,0){20}}
\put(62.5,10.5){+}
{\thicklines
\put(100,10){\line(-2,1){30}}
\put(100,10){\line(-2,-1){30}}
\put(80,10){\oval(5,20)}
\put(80,20){\circle*{1}}
\put(80,0){\circle*{1}}
\put(100,10){\line(1,1){10}}
\put(100,10){\line(1,-1){10}}
\put(100,10){\circle*{3}}}
\put(120,10){=}
\put(130,10){{\Large 0}}
\end{picture}

\bigskip
\bigskip
\bigskip
\bigskip
\bigskip
\bigskip
\bigskip
\bigskip
\bigskip
\bigskip
\bigskip
\bigskip

\begin{center}
Fig. 1.
\end{center}

\newpage
\begin{picture}(160,55)

{\thicklines
\put(5,40.5){\line(-1,1){5}}
\put(5,40.5){\circle*{3}}}
\put(5,40.5){\line(-1,-1){5}}
\put(5,40.9){\line(1,0){7}}
\put(5,40.1){\line(1,0){7}}
\put(17.5,40){=}
{\thicklines
\put(32.5,40.5){\line(-1,1){5}}
\put(42.5,40.5){\oval(20,10)[t]}
\put(52.5,40.9){\line(1,0){7}}
\put(32.5,40.5)
{\circle*{1}}
\put(52.5,40.5){\circle*{3}}}
\put(32.5,40.5){\line(-1,-1){5}}
\put(52.5,40.1){\line(1,0){7}}
\put(32.5,40.5){\line(1,0){20}}
\put(63,40){+}
{\thicklines
\put(100,40.5){\line(-2,1){30}}
\put(100,40.5){\line(-2,-1){30}}
\put(80,40.5){\oval(5,20)}
\put(80,50.5){\circle*{1}}
\put(80,30.5){\circle*{1}}
\put(100,40.9){\line(1,0){10}}
\put(100,40.1){\line(1,0){10}}
\put(100,40.5){\circle*{3}}}}
\put(120,40){+}
\put(0,0){+}
{\thicklines
\put(40,0.5){\line(-2,1){30}}
\put(40,0.5){\line(-2,-1){30}}
\multiput(20,10.5)(0,-2.2){9}%
{\circle*{1}}
\put(20,10.5){\circle*{1}}
\put(20,-9.5){\circle*{1}}
\put(40,0.9){\line(1,0){10}}
\put(40,0.1){\line(1,0){10}}
\put(40,0.5){\circle*{3}}}
\put(60,0){+}
{\thicklines
\put(100,0.5){\line(-2,1){30}}
\put(100,0.5){\line(-2,-1){30}}
\put(80.5,-9.5){\line(0,1){20}}
\put(79.5,-9.5){\line(0,1){20}}
\put(80,10.5){\circle*{1}}
\put(80,-9.5){\circle*{1}}
\put(100,0.9){\line(1,0){10}}
\put(100,0.1){\line(1,0){10}}
\put(100,0.5){\circle*{3}}
\end{picture}
\bigskip
\bigskip
\bigskip
\bigskip
\bigskip
\bigskip
\bigskip
\bigskip
\bigskip
\bigskip
\bigskip

\begin{center}
Fig. 2.
\end{center}

\end{document}